**Title: Perovskite Nanocrystals as Emerging Single-Photon Emitters: Progress, Challenges, and Opportunities**

Jehyeok Ryu[1,3], Victor Krivenkov[2,3], Adam Olejniczak[2], Alexey Y. Nikitin[1,4,a)], Yury Rakovich[1,2,3,4,a)]

[1]Donostia International Physics Center (DIPC), Donostia-San Sebastián 20018, Spain.

[2]Centro de Física de Materiales (CFM), Donostia - San Sebastián, 20018, Spain

[3]Polymers and Materials: Physics, Chemistry and Technology, Chemistry Faculty, University of the Basque Country (UPV/EHU), Donostia-San Sebastián 20018, Spain

[4]IKERBASQUE, Basque Foundation for Science, Bilbao 48013, Spain

a)Authors to whom correspondence should be addressed: alexey@dipc.org and yury.rakovich@ehu.eus

Abstract: Metal-halide perovskite nanocrystals (PNCs) have emerged as leading candidates for next-generation quantum emitters, offering a unique combination of high photoluminescence quantum yield, tunable emission, short radiative lifetimes, and record-high single-photon purity under ambient conditions. These properties, together with low-cost and scalable solution-phase fabrication, position PNCs as attractive alternatives to traditional epitaxial and colloidal quantum dots. In this Review, we outline the physical parameters that define quantum emission in PNCs, compare their performance to other established and emerging quantum emitters, and assess the key figures of merit, including photostability, single-photon purity, and photon indistinguishability, required for practical quantum applications. We discuss underlying mechanisms affecting PNC emission behavior and highlight recent advances in improving their quantum emitting properties through synthetic and photonic engineering approaches. While challenges related to environmental stability and photon indistinguishability remain, emerging strategies, such as surface passivation, metal ion doping, and coupling with electromagnetic nano- and microcavities, are steadily closing the gap between PNCs and ideal quantum light sources.

**Introduction**

Over the past two decades, secure quantum communication has become a major area of quantum information research due to its critical applications in finance, industry, and defense. Quantum communication broadly encompasses two key objectives: (1) connecting quantum computers to form quantum networks, and (2) enabling quantum key distribution (QKD) for secure information transfer. The fundamental unit of information in quantum communication is the qubit, a quantum system with no classical analog, defined as a coherent superposition of two quantum states. Photons are particularly suitable for qubit encoding for long-distance quantum communication because they weakly interact with the environment allowing, for example, free-space satellite–ground QKD distance up to 2000 km[1]. Classical QKD protocols (e.g., BB84) employ single-photon sources for generation of qubit encoded light. Light sources that emit single photons on demand with sub-Poissonian statistics are known as quantum emitters (QEs).

Perovskite nanocrystals (PNCs), a new class of colloidal semiconductor nanomaterials discovered in the past decade[2], which are able to emit photons as a result of the radiative recombination of quantum-confined exciton, have emerged as promising QEs. The ideal QE emits pure single photons on demand with a high repetition rate and 100% efficiency. No existing solid-state emitter perfectly meets all these criteria, but PNCs demonstrate several advantages for room-temperature operation compared to traditional QEs such as epitaxial and II-VI colloidal quantum dots (QDs) and color centers in diamond. In some respects, they even rival emerging two-dimensional emitters in hexagonal boron nitride or transition-metal dichalcogenides (Table I).

|  | PNCs | CQDs | EQDs | TMDC | hBN | Vacancies in diamonds |
|---|---|---|---|---|---|---|
| Emission wavelength | ~1.82-2.75 eV [3–5] | ~0.83-2.21 eV [6] | ~0.8-4.43 eV [7] | ~1.12-1.96 eV [8–14] | 1.55-4.13 eV [15–17] | NV- ~1.95 eV [18] (ZPL) SiV ~1.68 eV (ZPL) [19] GeV ~2.06 eV (ZPL) [20] |
| Photon absorption cross-section | >$10^{-13}$ cm² [21,22] | Up to $10^{-13}$ cm² (CdSe) [23] | $2.8 \times 10^{-14}$ cm² (InGaAs) [24] | - | - | $1.4–4.2\times 10^{-14}$ cm² (SiV-) [25] |
| Single-photon purity | CsPbBr$_3$: 0.02 (RT) [5] 0.019 (4 K, spectrally filtered) [4] | InP: 0.03 (RT) [26] ~0.08 (4 K) [27] | 0.23 (GaN, RT) [28] >$7.5 \times 10^{-5}$ (InGaAs, 4.2 K) [29] | WSe$_2$: 0.27 (150 K) [30], 0.02 (4 K) [31] | <0.2 (RT) [32] ~0.2 (4.5 K) [33] | SiV-: 0.1 (RT) [34,35], 0.11 (2K) [36] |
| Fluorescence lifetime | 5 ns (RT) [37] 186 ps (4 K) [4] | InP: 15 ns (4 K) [27], 21 ns (RT) [26] | InAs 1.8 ns (RT) [38] GaAs 100 ps (5 K) [39] | WSe$_2$ 4.1 ns (RT) [40] MoSe$_2$ 1.8 ps (7 K) [41] | 3.1 ns (RT) [42] 1.85 ns (70 K) [43] | SiV-: 1.28 ns (RT), 1.72 ns (4 K) [44] |
| PL QY | >0.95 (RT) [37] 0.956 (3.6 K) [45] | 1(CdSe 30-300 K) [46] | InGaAs: 0.11 (RT) [47] 0.85 (10 K)[48] | WSe$_2$ 0.6 (RT) [40] | 0.1-0.87 (RT) [49] ~0.1-0.2 (10 K) [50] | SiV-: 0.003-0.092 (RT) [25] SiV ~0.05 (77 K) [51] |
| Low temperature FWHM (T2) | 17 µeV (3.6 K) [45] | 5.2 µeV (InP, 4 K) [27] | 1.8 µeV (InGaAs, 2 K) [52] | 1.5 meV (MoSe$_2$, 1.8 K) [53] | 26.9 µeV (5 K) [54] | 1.6 µeV (SiV-, 6 K) [54] |
| HOM visibility | 0.56 (4 K) [4] | - | 0.99 (GaAs, 4K) [55] | - | 0.56 [56] | 0.8 (NV-, 4K) [57] |

**Table I.** Properties of cavity-free QEs

The most fundamental requirement for any QE is high single-photon purity, i.e. a very low probability of emitting two or more photons simultaneously. The standard metric is the second-order correlation function measured via a Hanbury Brown–Twiss (HBT) setup at zero time delay between two detectors, $g^{(2)}(0)$[58]. For an ideal single-photon emitter $g^{(2)}(0) = 0$, in practice, however, lower $g^{(2)}(0)$ values indicate higher single-photon purity. Among all known cavity-free QEs at room temperature, PNCs hold the record with $g^{(2)}(0) \approx 0.02$ (98% single-photon purity)[5], followed by core-shell II-VI colloidal quantum dots (CQDs) with $g^{(2)}(0) \approx 0.03$[26]. At cryogenic temperatures, InGaAs epitaxial quantum dots (EQDs) can achieve nearly ideal single-photon purity, with the best value of $g^{(2)}(0) \approx 7.5 \cdot 10^{-5}$ [29] at 4K, however, at room temperature for cavity free epitaxial QDs the best value is dramatically increasing to the level of $g^{(2)}(0) \approx 0.23$ reported for GaN EQD[28]. Thus, PNCs offer a clear advantage in single-photon purity at room temperature compared to other QEs (Table I).

Another critical figure of merit for a QE is its brightness, which can be quantified as photon emission rate at CW excitation. At low excitation intensity $I_{exc}$, this rate can be expressed as $f \sim \frac{\sigma I_{exc}}{\hbar \omega} \Phi_{PL}$, where $\sigma$ is the photon absorption cross-section, $\hbar \omega$ is the energy of the photon, and $\Phi_{PL}$ is the photoluminescence (PL) quantum yield (QY) [59]. In the photon absorption saturation regime, the emission rate is given by $f \sim \frac{1}{2\tau_{PL}} \Phi_{PL}$, where $\tau_{PL}$ is the fluorescence lifetime [59]. For on-demand operation, a QE must emit a photon for each excitation pulse. This requires both a high PL QY and a large dipole moment of the optical resonant transition [60], i.e. bigger photon absorption cross-sections. Among all QEs, PNCs exhibit the highest reported photon absorption cross-sections, with a maximum value of $6.85 \cdot 10^{-13}$ cm$^2$ [61,23]. The highest reported PL QY values for cavity-free QEs belong to II-VI CQDs and PNCs with PL QYs exceeding 95% at both room and cryogenic temperatures [37,45] (Table I). At the same time, PNCs with high PL QY exhibit shorter fluorescence lifetimes compared to II-VI CQDs, both at room and cryogenic temperatures (Table I). Thus, due to the combination of high PL QY and short fluorescence lifetime, PNCs can be considered as the optimal choice for on-demand operation.

Beyond single-photon purity and brightness, advanced applications such as quantum communication protocols, quantum teleportation and quantum sensing, demand the generation of quantum entangled photon pairs, which can be produced from indistinguishable photons[60]. This requires emission of consecutive single photons exhibiting mutual two-photon interference effects, ensuring full coherence. The primary parameter that quantifies the level of indistinguishability is the Hong-Ou-Mandel (HOM) visibility ($V_{HOM}$), which can be quantified by measuring the coincidence levels of indistinguishable ($C$) and distinguishable ($C'$) photons: $V_{HOM} = (C' - C)/C'$ [4]. For an ideal QE, this parameter is equal to 1. However, for real QEs, it is limited by dephasing processes occurring within the QE. The coherence time $T_2$ in the emitter can be expressed as $\frac{1}{T_2} = \frac{1}{2T_1} + \frac{1}{T_2^*}$, where $\frac{1}{2T_1}$ represents homogeneous broadening of the PL spectrum line due to spontaneous emission with a lifetime of $T_1$, and $T_2^*$ is the dephasing time, accounting for all other decoherence processes, induced by the interaction of the exciton with the environment and phonons. For an ideal QE, the relation simplifies to: $\frac{1}{T_2} = \frac{1}{2T_1}$, which is the Fourier transform limit. Thus, the goal is to reach $T_1 \ll T_2^*$ by lowering $T_1$ and elongating $T_2^*$. To date, the photon indistinguishability can be achieved only at cryogenic temperatures, where $T_2^*$ is long enough for practical applications. The best values achieved for cavity-free QEs are 0.99 for GaAs epitaxial QDs, 0.8 (NV⁻) [57], and 0.56 for both PNCs [4] and BN [56]. Conventional colloidal QDs can't achieve this limit due to long lifetime $T_1$ from the dark-state emission which become prominent at low temperatures [27]. Thus, with further improvements in their properties, PNCs hold strong potential to become reliable sources of indistinguishable single photons, providing an alternative to classical QEs.

Having presented and discussed the main parameters that allow PNCs to be considered as promising highly efficient QEs. Further, in this review we will discuss the fundamental mechanisms making PNCs promising candidates as a QE and current challenges for PNCs. We will also survey current strategies to overcome the remaining limitations of PNCs—namely, improving their stability and coherence—toward approaching the ideal QE.

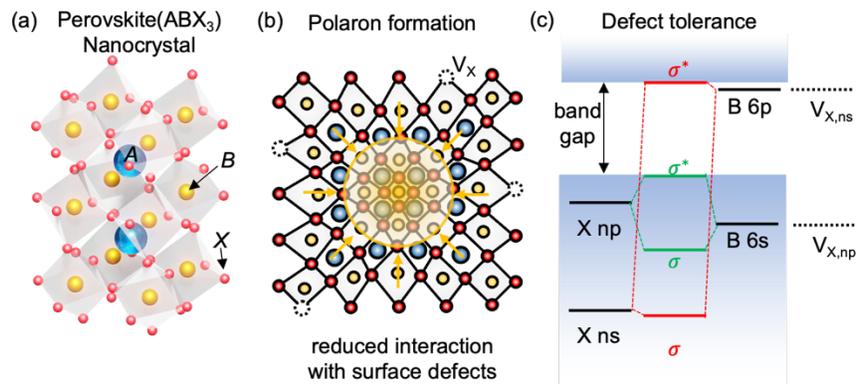

**Figure 1. Crystal structure and defect tolerance of perovskite.** (a) three-dimensional schematic illustration of the tilted corner-sharing octahedral crystal structure, which is a typical case of perovskite as tetragonal or orthorhombic phase. Blue, yellow, and red spheres represent A, B, and X atoms, respectively, while gray octahedron represent $BX_6$ structure. (b) Schematic of polaron formation inside a perovskite nanocrystal, illustrating how a localized charge carrier reduces the interaction from the surface trap sites. (c) Schematic energy-level diagram of the dominant vacancy defects in perovskite; $\sigma(\sigma^*)$ represents (anti)bonding states from the atomic orbitals.[62]

**Defect tolerance and Structural lability**

Metal-halide perovskites typically have the crystal formula $ABX_3$ as described in Figure. 1a, where A is a monovalent cation (e.g., Cs+, methylammonium MA+, or formamidinium FA+), B is a divalent metal cation (commonly Pb2+), and X is a halide anion (Cl−, Br−, or I−). The radii of the A and X ions affect the structural stability of the perovskite lattice, which is often quantified by the Goldschmidt tolerance factor (stability for values ~0.8–1.0) and the octahedral factor (optimal range ~0.4–0.6)[63]. Depending on temperature and composition, perovskites can adopt different phases (cubic, tetragonal, or orthorhombic) that are all derived from the ideal cubic structure. For example, all-inorganic $CsPbBr_3$ PNCs undergo phase transitions from cubic to tetragonal around 403 K, and from tetragonal to orthorhombic around 361 K as the temperature is further lowered[64]. Lead-halide perovskites also possess relatively soft lattices, prone to local distortions. Such dynamic lattice behavior does not only give rise to phase fluctuations and defects, but also provides a favorable environment for strong carrier-phonon interactions, which in turn promote the formation of polaron. The formation of polarons (charge carriers coupled with local lattice distortions) tends to localize charge carriers and screen them from scattering at defect sites (Figure 1b)[62,65–69]. In essence, polarons stabilize charge carriers by reducing their interaction with surface traps, which helps maintain high radiative recombination efficiency even in the presence of lattice imperfections.

In addition, the electronic band structure of lead-halide perovskites imparts a notable degree of defect tolerance to their optical properties. The valence band maximum is an antibonding combination of B 6s and X p orbitals (3p for Cl, 4p for Br, 5p for I), while the conduction band minimum is B 6p dominated and corresponds to the symmetry allowed antibonding B 6p, X s state (Figure 1c)[70,71]. A key consequence of this is the formation of shallow point defects. Because the band edges have antibonding character and are derived from different atomic parents, the dominant vacancies and interstitials introduce levels located close to the band edges rather than deep in the gap.[68] Therefore, a charge carrier trapped in these defect states can easily thermally escape at room temperature. Meanwhile, antisites defects can introduce deep trap sites, but they are energetically unfavorable due to their high formation energies. Together, these factors make PNCs "defect-tolerant" emitters, often achieving PL QY approaching unity despite the presence of structural defects. This defect tolerance is a distinctive advantage of PNCs, enabling consistently high brightness and efficiency without the elaborate surface passivation which required for traditional QDs[72], where deep dangling bond traps make a core-shell configuration essential for high PL quantum yield.

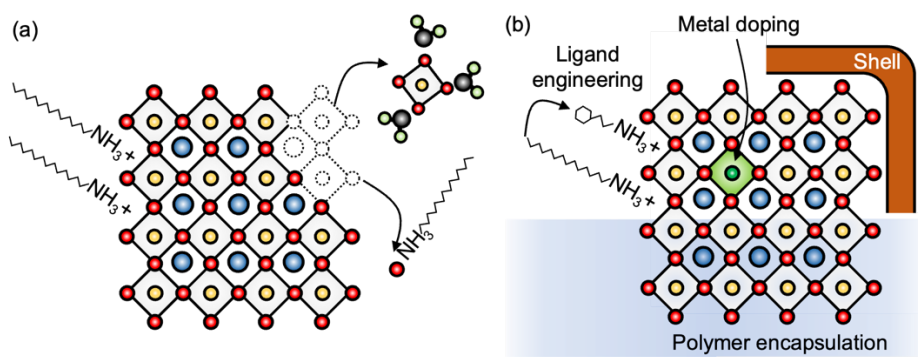

**Figure 2. Structural lability of perovskite nanocrystals and strategies to reinforce stability**. (a) Schematic illustration of the degradation pathway: highly dynamic ligands detach from the surface of the nanocrystal-removing surface Br⁻ and exposing undercoordinated sites-then $H_2O$ molecules (black and green circles) bind via coulombic interactions and disassemble the lattices. (b) Schematic of stabilization strategies: ligand engineering for stronger surface binding, metal-ion doping, core-shell architectures, and polymer encapsulation to strengthen surface binding, passivate defects, and prevent the surface to be exposed.

However, structural lability remains one of the main issues hindering PNCs widespread application. The soft ionic nature makes them highly susceptible to crystal structure degradation. Especially, the structure degradation is severe at the single particle level. Usually, in the presence of moisture and under light illumination, PNCs exhibit a continuous blueshift in their emission spectra before ultimately undergoing photobleaching as their size gradually shrinks due to the slow dissolution of PNCs in ambient moisture[73–75] as illustrated in Figure 2a. Therefore, several strategies have been introduced to enhance the stability of PNCs including surface ligands engineering, matrix encapsulation, inorganic shell formation, and modification of the PNC structure with metal doping as outlined below.

Ligands play a vital role during synthesis by preventing PNC from growing into the bulk phase and by maintaining colloidal stability[2]. Primary ligands used in colloidal QDs, oleic acid (OA) and oleylamine (OAm) were also widely used in PNCs synthesis. However, these highly dynamic, long-chain ligands can be easily detached from the PNC surface, leading to the formation of surface defects and increased susceptibility to degradation in the presence of ambient moisture (Figure 2a). This issue is particularly pronounced at single particle level, where highly diluted sample solution promote ligand desorption[76]. Therefore, different ligands were introduced to improve ligand binding to the PNCs (Figure 2b), by engineering either the "head" group (functional group that binds to the nanocrystal surface) or the "tail" group (lipophilic chain that stabilizes the PNC in nonpolar solvents). For instance,[73] Seth et al. demonstrated that post-synthetic treatment with alkylthiol ligands increases PL QY of PNCs by 1.4 times and improves their photostability at the single-particle level [77]. Bodnarchuk et al. reported that treatment with didodecyldimethylammonium bromide (DDAB) significantly enhanced PL QY from 60–70% to nearly 100%[78]. However, excessive surface passivation with bulky, long-chain ligands can destabilize PNCs due to intramolecular entropy reduction and increased surface free energy. For that reason, Mi et al. proposed an alternative approach using short-chain ligands terminated with an aromatic ring, such as phenethylammonium (PEA), which provides effective passivation via strong π-π interactions. This strategy notably reduced blinking in individual PNCs compared to those covered with DDA ligands [79]. Also zwitterionic molecules, which contain both positively and negatively charged groups simultaneously binding to the nanocrystal surface, were recently demonstrated as promising ligands for PNCs[80,81].

Encapsulating PNCs in a polymer matrix offers another effective strategy to improve their stability by isolating from the interaction with oxygen and water molecules (Figure 2b). This approach is particularly beneficial in single-particle spectroscopy, where exposure to ambient conditions can significantly alter PNCs properties. Raino et al. studied the influence of polymer matrices on the spectral stability of CsPbBr$_3$ PNCs covered with OA, OAm, and DDAB, and found that the polystyrene is the most effective matrix, whereas commonly used polymethylmethacrylate (PMMA) has a nonoptimal affinity to the hydrophobic ligands, higher residual water content under ambient conditions, and lower protection abilities against environmental contaminants, extending emission stability beyond 100 seconds [74]. Similar results were reported by Yu et al., who compared polystyrene and PMMA matrices under both ambient and inert conditions using OA, OAm, and DDAB ligands[82]. To further enhance stability, Mi et al. encapsulated PNCs covered with PEA ligands between two glass coverslips in a nitrogen atmosphere, sealing them with UV-curing optical adhesives[79]. This method, combined with ligand engineering, achieved remarkable photostability (over 30 minutes) and non bleaching emission for 600 minutes. However, this preparation method is not suitable for device integration. We believe that further optimization of the combinations of polymer matrices and various ligands can significantly enhance the stability of single PNC QEs under various environmental conditions.

Core-shell nanostructures represent a well-established approach for improving stability, enhancing PL QY, and reducing blinking. This strategy has been extensively explored for II-VI CQDs[83]. To this moment, a few studies demonstrate an emission of shell-protected individual PNCs, including perovskite/semiconductor, perovskite/perovskite, and perovskite/SiO$_2$ structures[84,85]. Tang et al. demonstrated high chemical stability and nonblinking PL behavior in single CsPbBr$_3$/CdS core/shell QDs for over 450 seconds [86]. Raino et al. investigated PbBr-depleted shells in diluted CsPbBr$_3$ PNC solutions, revealing a significant reduction in exciton-phonon interactions, leading also to the sufficient narrowing of the emission spectra [87]. Zhang et al. introduced an innovative approach, synthesizing FAPb(I$_{1-x}$Br$_x$)$_3$ perovskite QDs within an FAPbBr$_3$ matrix [88]. These buried PNCs exhibited nonblinking emission stability for 4000 seconds under 1 W·cm$^{-2}$. However, further improvements are still needed to optimize shell synthesis protocols and overcome challenges associated with the inherent soft ionic nature of perovskite structures and the lack of ideally tailored shells with near-perfect lattice matching.

Recently doping PNCs with transition metal ions has emerged as an effective strategy to enhance stability for applications in photovoltaics and lighting. Common dopants include nickel[89,90], tin[91], and zinc[92,93]. While most studies focus on stability in solution, D'Amato et al. investigated Zn-doped CsPbBr$_3$ PNCs at the single-particle level[76]. They reported enhanced spectral stability exceeding 60 minutes, with a minimal emission shift of < 4 nm. For undoped samples in ambient condition it is typical to observe the shift of over 10 nm during the several tens of seconds[74].

**Exciton Fine Structure and Dynamics**

Photoluminescence from PNCs originates from excitonic recombination, and it is dominated by the band-edge (lowest-energy) exciton due to ultrafast non-radiative relaxation, where carriers excited into higher energy levels quickly relax to the lowest exciton state, then emits a photon. The exciton fine structure – the subtle splitting of the energy level of the band-edge exciton into multiple sublevels – has attracted considerable interest because it governs the radiative dynamics at low temperature and affects photon coherence. As illustrated in Figure 3a, the energy level of the band-edge exciton in a PNC is split by the electron-hole exchange interaction into a lower-energy "dark" singlet state (total angular momentum $J = 0$, spin-forbidden optical transition) and a higher-energy "bright" triplet manifold ($J = 1$, spin-allowed transitions). The magnitude of the bright–dark splitting in energy increases as the PNC size decreases, since stronger quantum confinement enhances the electron-hole exchange interaction [94–96]. In a perfectly cubic lattice (high symmetry), the three triplet substates are degenerated. However, any deviation from cubic symmetry (for instance, the intrinsic lattice distortion in tetragonal or orthorhombic phases) lifts this degeneracy. Crystal field effects can further split the energy of the bright triplet into two sublevels in a tetragonal phase and into three sublevels in an orthorhombic phase [96,97]. These various influences on the exciton fine structure are depicted schematically in Figure 3a.

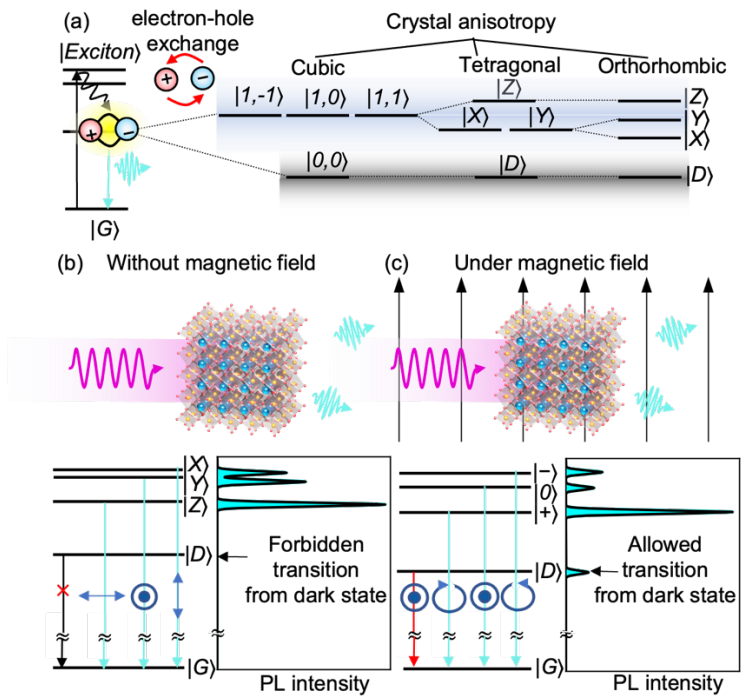

**Figure 3. Exciton fine structure of perovskite nanocrystals.** In the left side of the panel (a) display energy level diagram showing non-resonant optical pumping into high-lying excitonic states, ultrafast non-radiative relaxation into the lowest exciton manifold, and radiative recombination back to the ground state. The right side of the panel (a) describes the fine-structure splitting of that lowest exciton energy level: electron-hole exchange interactions separate it into a dark singlet ($|D\rangle$) and a bright triplet ($|X\rangle, |Y\rangle, |Z\rangle$), and crystal-field anisotropy further lifts the degeneracy of the triplet into three orthogonally polarized linear states. (b) Ordering and polarization selection rules for the exciton fine structure levels: at zero magnetic field, emission occurs only from the three linearly polarized bright states, while dark singlet remains optically inactive. Under an applied magnetic field, mixing bright and dark states activates the $|D\rangle$ state and produces circularly polarized emissions ($|-\rangle, |0\rangle, |+\rangle$) from bright triplet states.

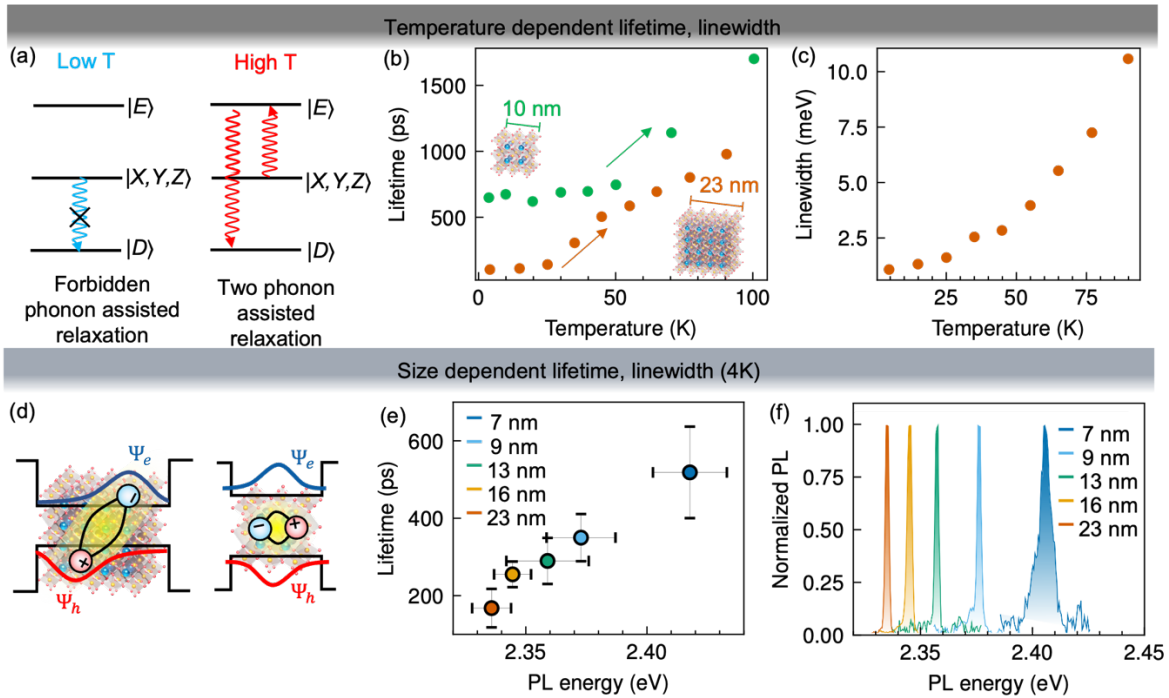

**Figure 4. Temperature and size dependence of exciton dynamics in perovskite nanocrystals.** (a) Energy level schematic showing relaxation pathways between the bright triplet states ($|X,Y,Z\rangle$), the dark singlet ($|D\rangle$), and a higher excited state ($|E\rangle$). At low temperature, the direct phonon assisted transition from the bright triplet states to the dark state is forbidden (left), so the bright states remain occupied dominantly. At higher temperature, two phonon assisted relaxation process enable excitation to $|E\rangle$ followed by relaxation into $|D\rangle$. (b) Temperature dependence of exciton lifetime for weakly confined regime (orange dots[98]) and intermediate confined regimes (green dots[99]), with both exhibiting shorter lifetimes at lower temperatures. (c) Temperature dependence of emission linewidth, narrowing at low temperature due to reduced electron-phonon dephasing. (d) Schematic of electron-hole wavefunction in larger (left) and smaller (right) nanocrystal; larger nanocrystal has stronger oscillator strength. (e) Size dependent exciton lifetime measured at 4K, showing shorter lifetimes for larger nanocrystals consistent with their stronger oscillator strength as illustrated in (d). (f) Size dependent emission linewidth, where smaller nanocrystals display broader linewidths due to stronger electron-phonon dephasing from a higher surface to volume ratio. (b, c, e, f) Zhu et al., Nature. **626**, 535 (2024); licensed under a Creative Commons Attribution (CC BY) license.

An important consequence of the exciton fine structure is the existence of a dark singlet state which exhibits spin-forbidden transition to the ground state. In conventional II–VI and III–V semiconductor QDs (e.g., CdSe or InGaAs dots), the dark exciton level lies below the bright exciton, and at cryogenic temperatures the exciton can relax into the dark state, resulting in weak or delayed emission (long-lived phosphorescence-like decay) [100]. Early studies of PNCs noted that they exhibit bright emission even at low temperatures despite the presence of the dark singlet state, leading to a debate as to whether the bright triplet could, in fact, be the lowest exciton state in these materials [94]. Some works hypothesized that a Rashba-type energy level splitting (arising from structural inversion asymmetry) might invert the level ordering and place the energy of a bright exciton state below the one of the dark state [94–96]. However, subsequent magneto-optical experiments conclusively showed that the dark singlet remains the true ground state. In FAPbBr$_3$ and

CsPbI$_3$ perovskite crystals, the nominally "dark" exciton transition can be activated and observed under high magnetic fields, confirming its identity as the lowest exciton level (Figure 3b,c) [101–103]. The current understanding is that two phonon-assisted relaxation of the exciton is required for it to populate the dark state from the bright state [101,102] as described in Figure 4a. At very low temperatures, this phonon-mediated transition becomes inefficient – particularly when the bright–dark energy splitting is large – forcing the exciton to remain in a bright state and decay radiatively. Consequently, PNCs can sustain predominantly bright exciton emission even at cryogenic temperatures. This behavior stands in stark contrast to classical CQDs, which have a dark ground exciton that causes their emission to become dim and slow at low temperature [100]. This exceptional ability of PNCs to maintain fast, bright radiative decay at cryogenic temperatures positions them as a superior candidate for next-generation quantum light sources, where both coherence and high brightness are critical.

As the temperature increases, thermally activated phonon processes begin to populate the dark state of an exciton, slowing down the radiative decay rate (Figure 4b). Thus, cryogenic temperatures are required to keep excitons in the bright state, preserving a fast radiative decay and bright emission. Even in the regime where the dark state is largely unoccupied (e.g., at liquid helium temperatures), the fluorescence lifetime of the bright exciton in PNCs is strongly size-dependent. Larger PNCs generally exhibit shorter fluorescence lifetimes reflecting the higher oscillator strength of the exciton than the one of smaller PNCs, an effect often described as the "giant oscillator strength" phenomenon [94,104]. This arises because in larger PNCs, the electron and hole wavefunctions are more spatially delocalized and overlapping, having electron-hole correlated motion, which results in a larger transition dipole moment. Conversely, strong confinement in smaller PNCs reduces the electron–hole correlated motion and therefore reduced oscillator strength of the optical transition from an exciton as schematically described in Figure 4d. The trend is evident in size dependent lifetime measurements (Figure 4e): smaller PNCs show longer exciton lifetimes, whereas larger PNCs have significantly faster radiative decay of the exciton under the same conditions (once the dark state is unoccupied). Moreover, the temperature at which the bright exciton dominates over the dark exciton depends on the bright–dark energy splitting as a consequence on the size of PNCs. Smaller PNCs, with larger exciton energy level splitting due to stronger electron-hole exchange interaction, can maintain predominantly bright-exciton emission up to higher temperatures (observed up to ~50 K) [99], whereas larger PNCs (with smaller energy level splitting) transition to predominantly dark-exciton emission at lower temperatures (~30 K) [98]. Above these thresholds, the effective PL lifetime lengthens for all sizes due to the increasing contribution of the dark state as a non-radiative reservoir.

Thermal effects also impact the emission linewidth and coherence. As temperature rises, the homogeneous linewidth broadens due to stronger exciton–phonon coupling: interactions with phonons cause rapid dephasing, leading to a monotonic increase of the emission linewidth with temperature (Figure 4c). Environmental interactions (such as fluctuations in the local electrostatic or strain environment) can further broaden the optical transitions. Generally, smaller PNCs – with their larger surface-to-volume ratio – are more susceptible to environmental perturbations such as (among others) surface charges or vibrations from polymer encapsulation layer, and thus often exhibit broader emission linewidths or spectral diffusion compared to larger NCs. Considering these factors together, it becomes clear that larger PNCs operated at low temperature offer significant advantages for generating highly-coherent single photons: they combine fast radiative rates with relatively narrow linewidths. On the other hand, smaller PNCs offer the benefit of a larger bright–dark energy level splitting, which helps sustain emission from the bright states up to higher temperatures. These trade-offs must be considered when optimizing PNCs for generating coherent single photon emission, and they point toward controlling PNC size and composition as a means to tailor emission properties for specific operating conditions.

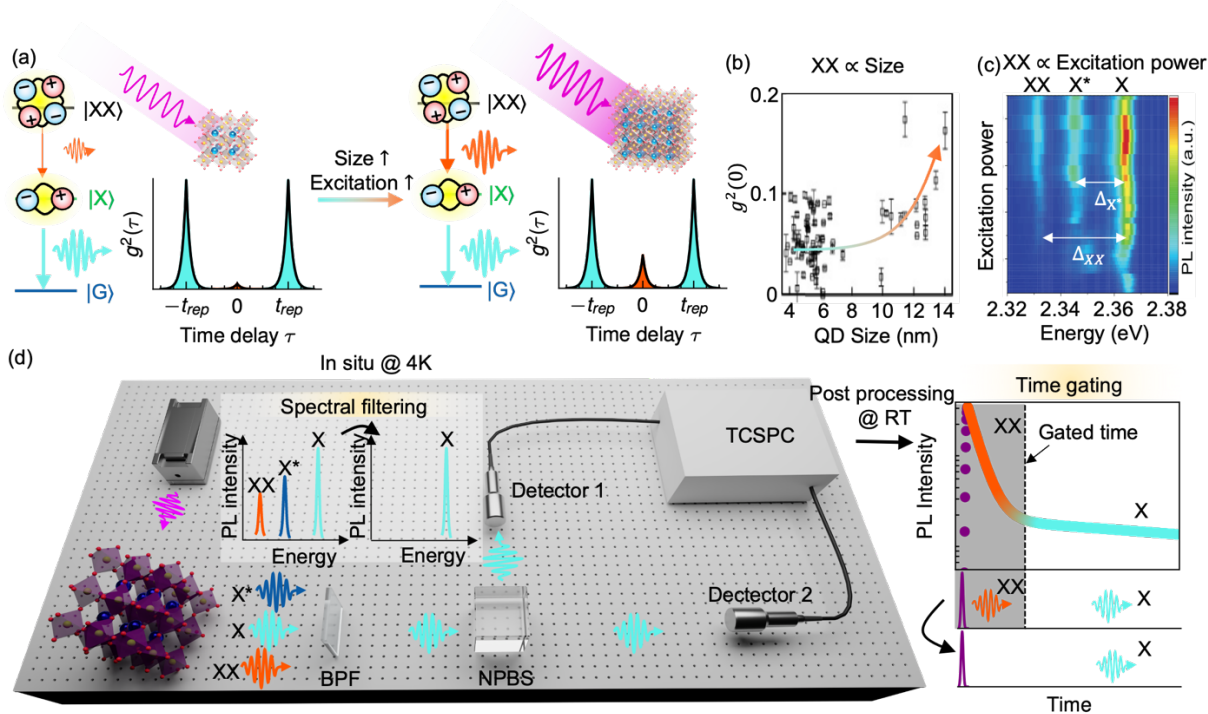

**Figure 5. Exciton complexes and single photon purity of perovskite nanocrystals.** (a) Schematic of biexciton-exciton cascade of smaller (left) and larger nanocrystals (right). Increased volume or higher excitation fluence raises the biexciton generation probability, which lower (increase) single photon purity ($g^{(2)}(0)$) as also described in (b) Mi et al., Nat.Comm. **16**, 204 (2025); licensed under a Creative Commons Attribution (CC BY) license.[79] and (c) Zhu et al., Adv.Mater. **35**, 2208354 (2023); licensed under a Creative Commons Attribution (CC BY) license[105]. (d) Schematic of two purity-enhancement techniques: spectral filtering of the exciton emission in a Hanbury-Brown and Twiss set up (left) and post-processed time gating of photon arrival histogram (right).

**Single photon purity**

Under certain conditions more than one exciton can be generated either in larger size semiconductor nanocrystals and PNCs or at high excitation intensities [106,107]. Due to the strong Coulomb interaction between excitons, these excitons can form multiexciton complexes that consist of two or more electron-hole pairs bound to each other. This enables the possibility of the cascade recombination of the multiexcitons followed by the emission of several photons, thus increasing the chance for the non-single-photon emission from PNCs. However, the Coulomb interaction also leads to a high rate of nonradiative Auger recombination of the multiexcitons, in which the energy from a recombining electron-hole pair may not be emitted as a photon but instead transferred to a third carrier (either an electron or a hole from the other excitons) [108]. This carrier is then excited to a higher energy state and subsequently relaxes back to its initial state via thermalization. The more electron-hole pairs a multiexciton complex consists of, the higher the probability of Auger recombination, because additional carriers are available to absorb the recombination energy. Consequently, among multiexciton complexes, the biexciton – comprising two electron-hole pairs –is most likely to emit photons after the single exciton. Therefore, we will focus now on how single photon purity of PNC emission depends on biexciton dynamics.

The biexciton-exciton-ground state cascade recombination results in the emission of two photons as illustrated in Figure 5a(left), and this process significantly affects the single-photon purity of PNC QEs. If the emission signal from the semiconductor QD is measured using HBT setup at low excitation intensity (average exciton occupancy $\ll$ 1) the value of the second order correlation function at zero delay of the two consecutive photon detection, $g^{(2)}(0)$ is related to the ratio of biexciton PL QY ($QY_{XX}$) to single exciton PL QY ($QY_X$) [109]:

$$g^{(2)}(0) = \frac{QY_{XX}}{QY_X}$$

Hence, achieving high single-photon purity requires maximizing the PL QY of the exciton transition while minimizing that of the biexciton. The purity of single-photon emission from PNCs is strongly size-dependent, with smaller PNCs generally exhibiting weaker biexciton emission than larger ones of the same material (Figure 5b) [5]. This is because the stronger quantum confinement in smaller PNCs enhances Coulomb interaction among carriers, leading to faster nonradiative Auger recombination. Consequently, one approach to improving single-photon purity of the emission is the synthesis of highly confined PNCs. However, reducing PNC size also results in a blueshift of the emission due to the quantum confinement effect, a decrease in absorption cross-section (which scales with volume) [61,110,111], and an increased surface-to-volume ratio affecting their stability. Thus, the choice of the PNC size should be carefully considered to achieve single-photon purity, taking into all these parameters.

The probability of biexciton generation increases with excitation power density (Figure 5c) [105]. Thus, one way to ensure pure single-photon emission is to use low-power excitation. However, since exciton emission intensity also scales with excitation power, higher excitation powers are required to achieve a high photon emission rate. Ideally, a perfect QE should operate at the saturation power level, where the emitter is being excited as frequently as possible, each excitation pulse generates a single photon, and the emission rate approaches its maximum. Even for highly confined QDs, biexciton emission may still become significant at high excitation densities. Therefore, filtering out biexciton photons is necessary to achieve highly pure single-photon emission. Two main methods may be used for this purpose: spectral filtering and temporal filtering (time gating) as described in Figure 5d.

The photon emitted during the biexciton-to-exciton transition has a lower energy ($E_{XX}$) than the photon emitted during the exciton-to-ground-state transition ($E_X$) by an amount equal to the biexciton binding energy ($\Delta_{XX}$), which is typically up to -40 meV (Figure 5c). As a result, biexciton emission is redshifted relative to exciton emission [98,105,112–114].

This energy difference is too small to be resolved at room temperature, where even single-particle emission spectra are broad. However, at sufficiently low temperatures (~4 K), exciton and biexciton emission lines become spectrally distinct, enabling spectral filtering using bandpass filters placed in the optical path between the emitter and the beam splitter (Figure 5d, left). This method allows photons emitted by biexcitons to be filtered out during the experiment. However, due to the small spectral separation, optical filters of high precision are required, and this technique is mostly applicable at cryogenic temperatures [4,45,98].

An alternative approach is temporal filtering, or time gating, which can be applied at any temperature and in post-processing of already detected photons. In time-correlated single-photon counting (TCSPC) method, the arrival time of each detected photon relative to the preceding laser pulse is precisely recorded. Since biexciton emission occurs on a much shorter timescale than exciton emission, selecting only photons detected after a chosen delay during post-processing can completely eliminate biexciton contributions, achieving $g^{(2)}(0) = 0$ (Figure 5d, right) [115].

**Coherent Single-Photon Emission**

For quantum interference experiments and quantum optical computing, the coherence and indistinguishability of emitted photons are as important as the single-photon purity. To obtain highly indistinguishable photons, the emitter's optical coherence time must be comparable to twice its PL lifetime: $T_2 \approx 2 \cdot T_1$ to reach the Fourier transform limit, which possesses maximal temporal coherence. Figure 6a illustrates this concept: if an emitter undergoes radiative recombination (emits a photon) before any dephasing event occurs, then the emitted photon retains a well-defined phase relation to other photons. In contrast, if a dephasing event (loss of phase coherence due to interactions with the environment) happens on a timescale shorter than the emission, the photon's coherence is compromised. Hence, coherent single-photon emission requires that spontaneous emission be faster than dephasing. The ideal scenario is the above mentioned Fourier-limited case, where the emission linewidth is determined solely by the excited-state lifetime and not broadened further by dephasing.

In PNCs, substantial progress has been made towards improving photon coherence at low temperatures. Experiments on relatively large $CsPbBr_3$ PNCs (which, as noted, have high oscillator strengths and low

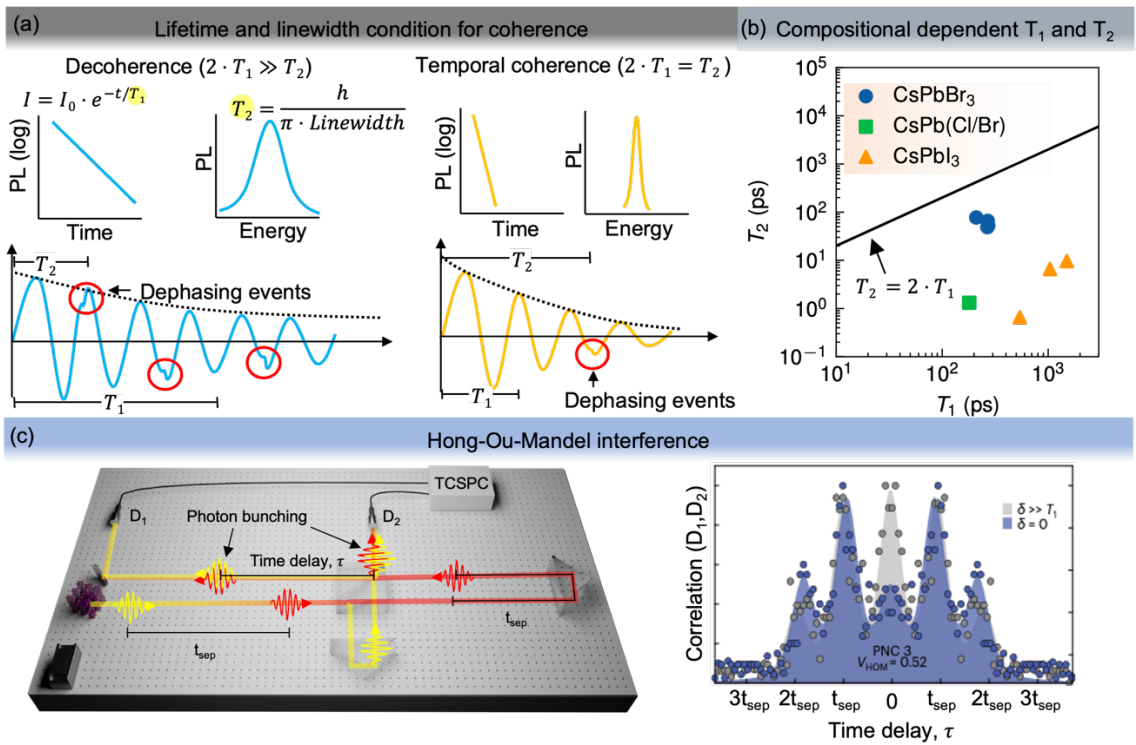

**Figure 6. Coherence and indistinguishability of single photons from perovskite nanocrystals.** (a) Temporal emission schematics contrasting decoherent versus coherent radiative decay of exciton. In the decoherent case (left), exciton dephasing events ($T_2$) occur before radiative decay ($T_1$); in the coherent case (right), the exciton decays radiatively prior to any dephasing. (b) Reported fluorescence lifetime ($T_1$) and coherence times ($T_2$) for various perovskite nanocrystals. (c) Hong-Ou-Mandel interferometer set up (left) used to assess photon indistinguishability, alongside the measured two-photon interference visibility for $CsPbBr_3$ nanocrystal (right). Reproduced with permission from Nat.Phot. **17**, 775 (2023). Copyright 2023 Springer Nature.

inhomogeneous broadening) have shown emission linewidths approaching the Fourier-transform limit at cryogenic temperatures (Figure 6b). Consistent with these findings, two-photon interference measurements have confirmed a degree of indistinguishability in photons from single PNCs as the experimental set up is schematically illustrated in the Figure 6c. For example, Hong–Ou–Mandel interference experiments on single $CsPbBr_3$ PNCs have reported HOM visibilities $V_{HOM} > 0.5$ [4]. While the reported visibility value may seem modest in absolute terms, it is a remarkable achievement for a colloidal emitter operating in the visible range, and it unambiguously demonstrates the presence of quantum interference (indistinguishability) between photons emitted by PNCs. By comparison, the best HOM interference results from EQDs have

$V_{HOM} > 0.99$ under similar conditions [55]. Clearly, there is still a gap to close for PNCs to reach that level of performance. Nonetheless, these initial demonstrations establish PNCs as viable sources of partially coherent photons. In addition to HOM tests, other advanced techniques such as photon-correlation Fourier spectroscopy (PCFS) have been used to characterize the coherence of PNC emission spectra [45,116], and even collective coherent phenomena like superfluorescence have been observed in dense ensembles of PNCs under pulsed excitation, indicating that multiple excitons can phase-lock and emit cooperatively in phase [104,117,118]. These studies underscore the growing understanding and control of coherence in perovskite QEs.

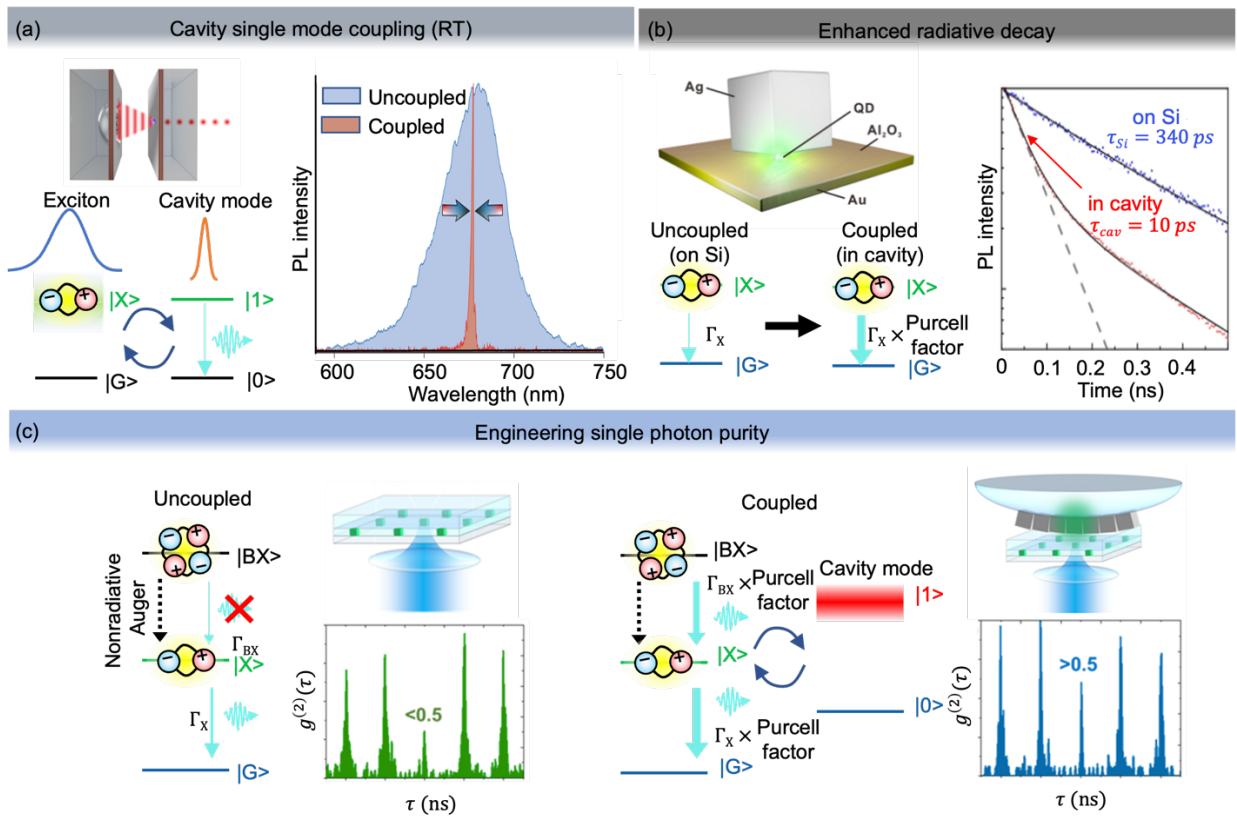

**Figure 7. Effects of light–matter coupling on the emission properties of individual PNC QEs.** (a) Narrowing of the emission spectrum of an individual PNC due to the coupling with an optical microcavity mode. Farrow et al., Nano.Lett. **23**, 10667 (2023); licensed under a Creative Commons Attribution (CC BY) license. (b) Shortening of the PL lifetime of a single $CsPbBr_3$ PNC due to coupling to a plasmonic nanogap cavity (silver nanocube on the gold mirror). Adapted with permission from ACS.Nano. **14**, 11670 (2020). Copyright 2020 American Chemical Society.[119] (c) Switch from the single-photon emission to the multiphoton emission of an individual PNC due to increased PL efficiency of the biexciton transition, induced by coupling to a plasmonic metasurface. Olejniczak et al., APL.Phot. **9**, 016107 (2024); licensed under a Creative Commons Attribution (CC BY) license [120]

**Coupling with a cavity**

Coupling colloidal quantum emitters, such as perovskite nanocrystals and quantum dots, with electromagnetic modes of micro- and nanocavities is one of the most effective ways to improve their properties as QEs [6]. The coupling between exciton states of PNC and an electromagnetic mode of the cavity also known as the light-matter coupling phenomenon can provide several significant advantages for PNC-based QEs. The main parameter for the exciton-cavity coupling is the coupling strength ($g$), which can be calculated using the following equation [6]:

$$g = \sqrt{N}\mu_{ij}\sqrt{\frac{\hbar\omega_0}{2\varepsilon\varepsilon_0 V_{cav}}}$$

Where $N$ is the number of equal oscillators involved in the interaction with the selected cavity mode, $\mu_{ij}$ is the dipole moment of the exciton transition, $\omega_0$ is a resonance frequency (equal for the exciton transition and cavity mode), and $V_{cav}$ is the cavity mode volume. For all single QE applications, $N = 1$. Depending on the ratio between the coupling strength and the damping rates of the cavity mode ($\Gamma_{cav}$) and the exciton transition ($\Gamma_X$), two main regimes of light-matter coupling are possible: weak light-matter coupling ($2g < \frac{\Gamma_{cav}+\Gamma_X}{2}$) and strong light-matter coupling ($2g > \frac{\Gamma_{cav}+\Gamma_X}{2}$) [121]. To date, strong coupling has been achieved for single colloidal QDs [122,123] but not for single PNCs. Therefore, in this section, we will focus on the weak coupling regime, which was recently demonstrated for PNCs using both optical [124] and plasmonic [120] cavities (the latter typically represented by metallic nanoparticles). The primary effect of the weak coupling regime is an increase in the density of photonic states at resonance frequencies (wavelengths) of the cavities. This leads to two practically important effects. First, if the light source used for the photoexcitation of QEs has a wavelength resonant with the cavity mode, the effective photoexcitation of the QE can be enhanced due to the local amplification of the electromagnetic field amplitude. This effect can lead to an increase in the photon absorption efficiency and to the corresponding increase of the QE PL emission intensity. The second effect is the Purcell enhancement, which occurs when the emission wavelength of the QE is resonant with the cavity mode. In this case, the local increase in the density of photonic states enhances the depopulation rate of the excited state of the QE, leading to the energy transfer from QE to the corresponding electromagnetic mode of the cavity. This energy can then be emitted as a photon or dissipated due to internal cavity losses, shortening the effective PL lifetime of the coupled emitter, which means shortening of $T_1$.

The main parameter for the Purcell effect is the Purcell factor ($F_p$), which is the ratio of the emission rate of the exciton coupled to the cavity ($\Gamma_X^{coupled}$) to the initial exciton emission rate ($\Gamma_X$). In the case of the resonant coupling of the emission wavelength with the cavity mode, it can be estimated by following equation:

$$F_p = \frac{\Gamma_X^{coupled}}{\Gamma_X} = \frac{3}{4\pi^2}\left(\frac{\lambda}{n}\right)^3\frac{Q}{V_{cav}}$$

where $\lambda$ is the emission wavelength (in vacuum), n is the refractive index of the media, $Q$ is the quality factor of the cavity, and $V_{cav}$ is the cavity mode volume. In practical implementations, nanophotonic structures like dielectric Mie cavities, photonic crystal cavities, or plasmonic nanoantennas are designed to achieve ultra-small mode volumes and high quality factors, thereby maximizing the Purcell enhancement.

The first demonstration of the Purcell effect for an ensemble of CsPbBr$_3$ PNCs coupled with photonic nanocavities was achieved in a photonic crystal cavity by Yang et al. [125]. The authors also noted that their system operated in the mode they called 'bad emitter' regime, where the emitter's linewidth is much broader than the cavity linewidth. In this limit, the Purcell factor becomes independent of the cavity quality factor and depends only on the emitter linewidth. They achieved a 2.9-fold enhancement in the average spontaneous emission rate, along with spectral narrowing and a 10-fold increase in brightness. This enhancement was obtained using a structure with a quality factor of 4000 and a calculated $V_{cav}$ of 0.55 $\left(\frac{\lambda}{n}\right)^3$. Similar Purcell factor values of 2–3 were achieved for a CsPbBr$_3$ PNC ensemble coupled to an array of Mie resonators (square silicon nanopillars), resulting in an 18-fold increase in emission intensity [126], as well as for a CsPb(Br$_x$Cl$_{1-x}$)$_3$ PNC ensemble coupled with SiO$_2$ microspheres [127]. The most significant enhancement among optical structures was observed for an ensemble of CsPbBr$_3$ PNCs interacting with a circular Bragg grating (also known as a bullseye cavity), which led to an order-of-magnitude increase in emission intensity and an 8-fold Purcell factor [128]. Additionally, this type of cavity was shown to enhance the directionality of PNC emission [129].

The first demonstration of the optical microcavity coupling of single CsPbI$_3$ PNC was achieved by Farrow et al. [124]. In their work the authors used a tunable open-access optical microcavity with $Q = 300$ and $V_{cav}$ of 0.5 µm$^3$, expecting a $F_P$ of 4.7. They also achieved single-photon purity of 0.06, blinking suppression, and spectral narrowing down to 1.1 nm full width at half maximum (FWHM) (Figure 7a). The narrowing of the emission spectrum due to the coupling with narrower cavity mode is expected as the alternative way to improve coherence. This effect could be used for narrowing the emission spectrum and emission wavelength selection. However further studies are necessary to understand the effect of that spectral narrowing on the T$_2$ parameter of the emitter.

Using plasmonic nanocavities instead of optical microcavities offers several advantages for light-matter coupling effects due to the ultrasmall mode volumes provided by plasmons [6]. In the work of Hsieh et al., a nanogap plasmon mode formed between a silver nanocube and a gold mirror enabled an extremely small $V_{cav}$ of approximately 0.002 $\lambda^3$. The authors reported a PNC lasing peak with a FWHM of 1.4 nm, corresponding to a $Q$ of ~382 [119]. Authors achieved lasing with a significantly shortened lifetime, decreasing from 340 ps on a Si substrate to just 10 ps inside the gap cavity (Figure 7b), which possibly can be attributed to the enhancement of the radiative rate of the exciton transition due to the Purcell effect as it is illustrated in the Figure 7b. Although the authors expected only a single PNC to couple with the gap mode due to the low density of PNCs on the sample surface, there was no clear evidence confirming single-emitter behavior or single-photon emission in the system.

Recently, Olejniczak et al. directly demonstrated reversible plasmon-exciton coupling for a single $CsPbBr_{2.5}I_{0.5}$ perovskite nanocrystal quantum emitter interacting with a metasurface composed of silver nanocubes [120]. The designed metasurface provided good spectral overlap between the bright scattering mode of the plasmonic metasurface and the emission transition of the PNC. The authors calculated a radiative rate acceleration factor of up to 12, which was attributed to the Purcell effect. This result significantly overcame the previously achieved enhancement for a $CsPbBr_3$ ensemble coupled with an Ag nanodisk array, where only a 1.65-fold enhancement of the PL relaxation rate was reported [130]. However, the strong radiative rate enhancement also led to an increase in the emission efficiency of the biexciton-exciton transition, which in turn reduced the single-photon purity (increase of the $g^{(2)}$ value) of the PNC QE (Figure 7c) [120]. Indeed, since exciton and biexciton emission spectra are very close in wavelengths, coupling with a broad plasmon mode accelerates both the biexciton and exciton radiative transitions, turning a non-emitting biexciton transition into the emitting one as illustrated in the Figure 7c. This can alter the ratio between the radiative and nonradiative rates of the biexciton transition, leading to an increase in the emission efficiency. Consequently, this raises the biexciton-to-exciton quantum yield ratio, resulting in a higher $g^{(2)}(0)$ value and a corresponding decrease in single-photon purity. This experiment pave a way to engineer multi-photons emission process.

Thus, both optical microcavities and plasmonic nanocavities can enhance the emission rate of PNC QEs and increase their emission intensity. This not only improves the operation rate and overall brightness of PNC QEs but also enables $T_1$ shortening, which is essential for reaching the Fourier transform limit. Optical microcavities, with their higher quality factors, are particularly useful for coupling with PNCs at ultralow temperatures. In that case, the spectral narrowing at low temperatures helping to avoid the situation where the emitter's linewidth is much broader than the cavity linewidth, and the Purcell factor becomes independent of the cavity quality factor [125]. Additionally, they can be employed for emission wavelength selection, spectral narrowing, and avoiding undesired coupling with biexciton transitions. On the other hand, plasmonic nanocavities offer higher Purcell factor values, demonstrating the potential for strong lifetime shortening at both low and room temperatures. However, precise coupling engineering is essential to suppress biexciton emission, thereby preserving the single-photon purity of PNC QEs.

**Outlook**

PNCs have rapidly advanced to the forefront of single-photon source research, combining attributes of traditional semiconductor QDs (high quantum yield, size-tunable emission) with unique advantages such as defect tolerance and fast radiative rates. As detailed in this review, PNCs can already meet or exceed the single-photon purity of other sources at room temperature, with $g^{(2)}(0)$ as low as 0.02 in the best cases, and partial optical coherence has been demonstrated at cryogenic temperatures. In the previous section, we discussed how to improve the optical coherence, accelerating spontaneous emission with a cavity so that radiative decay outpaces dephasing events. While this approach is effective, its practical liabilities are not negligible. For example, high Purcell factors typically demand ultrasmall mode volumes that make deterministic emitter placement difficult. Plasmonic resonators introduce ohmic losses that can drain radiative efficiency and distort far-field emission, and dielectric cavities that preserve quantum efficiency often struggle to surpass moderate Purcell factor values. This outlook therefore pivots to the physics of dephasing in PNCs and the materials/phononics strategies that can suppress it.

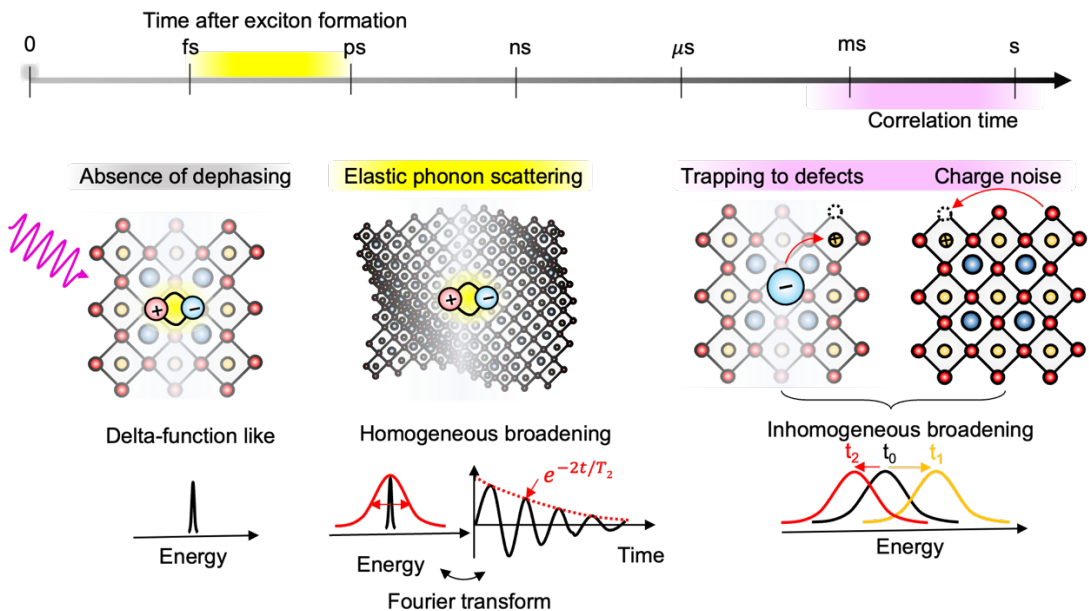

Figure 8. Schematic illustration of coherence loss mechanisms. Elastic phonon scattering contributes to the homogeneous broadening of the emission linewidth (middle, yellow). Inhomogeneous broadening of the linewidth kills the coherence immediately (right, pink)

A mechanistic chronology of coherence loss is sketched in Figure 8. Immediately after an exciton is formed in the band-edge state, it interacts with lattice vibrations. At cryogenic temperature, pure dephasing is dominated by elastic scattering with low-energy phonon modes[131]. This process scrambles phase and homogeneously broadens the zero-phonon line. Inelastic LO phonon scattering is suppressed at low temperature but becomes dominant as temperature rises, leading to the temperature dependent broadening seen in Figure 4c[98,101]. In parallel, inhomogeneous mechanisms act over $\mu s - s$ correlation times and manifests as spectral diffusion and intermittent spectral jumps (right panel of Figure 8). Intermittent trapping and recharging at defects toggle neutral excitons and trions, while mobile surface charges arising from dynamic ligands, halide vacancies, etc generate fluctuating internal fields. Crucially, both of inhomogeneous dephasing factors can be suppressed by stronger binding ligands with better surface passivation, and larger nanocrystals that reduce the charge noise from surface. Therefore, ligand engineering has advanced precisely to suppress these extrinsic channels. Conventional long-chain OA/OAm ligands are labile and undergo rapid binding-desorption equilibria, which promotes transient traps and charging. The first advance beyond OA/OAm was monocationic quaternary ammonium halides such as DDABr, which render the surface bromide rich and exchange resistant, improving robustness to purification and polar media[132]. The next advance was bidentate in effect through zwitterionic ligands that combine one quaternary ammonium head with one deprotonated anionic group to have an overall net-neutral charge; these dual moieties bind the lattice via opposite charges that the cationic ammonium stabilizing halide-terminated/A-site environments while the anionic group coordinates undercoordinated Pb–X motifs—yielding tight, low-desorption shells and suppressed intermittency[80,81]. Building directly on that dual-site logic, dicationic quaternary ammonium bromides, with two permanently charged N$^+$ heads linked by a short spacer, act as a functionally bidentate ionic anchor that can bridge neighboring A sites[116]; these bidentate ligands strongly suppresses inhomogenous broadening. In addition, π-stacking aromatic ammonium ligands phenethyl ammonium(PEA) create a compact, interlocking ligand layer whose intermolecular interactions lower the surface free energy and curb ligand loss; strongly confined $CsPbBr_3$ QDs with stacked PEA exhibit nearly non-blinking emission[26,77], while FAn, CFAn, and 3,4,5-FAn families exhibit stronger interfacial bonding than PEA but have thus far been explored primarily in perovskite solar-cell architectures[133,134]; translating these compact arylammoniums to colloidal nanocrystal shells remains an open and promising direction. Consequentially, the inhomogeneous broadening seems no longer issue. Correlation time resolved coherence time measurements on 20 nm $CsPbBr_3$ nanocrystals capped with zwitterionic ligands have found no change of coherence time from 10 μs – 5 ms scale without observable inhomogeneous dephasing factors, indicating elastic phonon scattering as the dominant dephasing factor[131]. In conclusion, recent advances in ligand engineering have improved optical coherence by effectively suppressing environmental fluctuations. Therefore, the remaining challenge for achieving idealistic optical coherence is to reduce pure dephasing caused by elastic phonon scattering. In the following paragraph, we will discuss a strategy to mitigate the elastic phonon scattering.

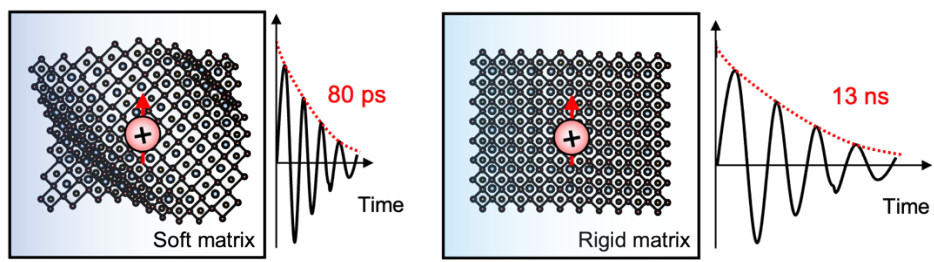

Figure 9. Increased spin coherence time of PNCs in rigid matrix. Schematic illustration of PNCs and corresponding spin coherence time (left: encapulsated in polymer matrix, right: embedded in glass matrix)

The practical consequences are straightforward. For typical bright PNC at low temperature is known to have the radiative relaxation time ($T_1$) ~ 100 – 200 ps[4,98] and the optical coherence time ($T_2$) ~ 50 – 80 ps[45], still not reaching the fourier transform limit. From an application perspective, a good cross-protocol benchmark is idealistic single-photon purity ($g^{(2)}(0) < 0.01$) and near-unity indistinguishability ($V_{HOM} > 0.99$) is required. Therefore, efficient suppression of elastic phonon scattering is required to meet the criteria. A potential strategy is to stiffen the vibrational environment using a rigid bath. The elastic dephasing rate effectively integrates the product of the exciton's deformation potential coupling and the strain spectral density. Increasing the effective elastic modulus, pushing confined acoustic modes to higher frequencies and introducing phononic gap can suppress this phonon density of states. For instance, color center in diamond has shown that $50-70$ GHz phononic bandgaps suppress resonant phonon channels and extend the orbital lifetimes by more than an order of magnitude, directly demonstrating that engineering the vibrational density of state can mitigate dephasing. In colloidal II-IV QDs, giant-shell CdSe/CdS and strain-graded core/shells reduce exciton-phonon coupling, suppress blinking, and stabilize spectra. Similarly, PNCs embedded in a rigid matrix can improve the coherence as seen in Figure 9. Ensembles of PNCs embedded in glass exhibit time-resolved Faraday ellipticity signatures of mode locking and nuclear assisted frequency focusing with hole spin coherence of 13 ns at 5 K[135], whereas similar ensemble of PNCs encapsulated in polymer matrix showed the hole spin coherence of 80 ps at 4.5 K[136]. Even though spin coherence is not directly related to the optical coherence, they are degraded by the same low energy phonon bath. Therefore, the fact that a rigid matrix improves orders of magnitude in spin coherence is clear evidence that it also effectively increase optical coherence time. The future direction could explore more facile way to prepare individual PNCs in a rigid matrix for optical coherence measurement.

In conclusion, metal-halide PNCs have established themselves as compelling single-photon emitters. Ongoing improvements in material and photonic engineering are rapidly closing the gap between PNCs and the "ideal" single-photon source. Given their fast progress and already impressive performance, it is foreseeable that PNC-based single-photon sources will play a significant role in next-generation quantum communication and optical computing platforms. The field is moving toward demonstrating PNC single-photon emitters that operate reliably at room temperature, with high purity, high brightness, and improving indistinguishability – a combination that could outperform many existing technologies. With continued interdisciplinary efforts in chemistry, materials science, and photonic engineering, PNC quantum emitters are on track to transition from laboratory curiosities to practical components of quantum technology.

**Acknowledgement**

The study was funded by the Department of Science, Universities and Innovation of the Basque Government (grants no. IT1526-22, PIBA_2024_1_0011, and PIBA-2023-1-0007) and the IKUR Strategy; by the Spanish Ministry of Science and Innovation (grants no. PID2022-141017OB-I00, TED2021-129457B-I00, PID2023-146442NB-I00, PID2023-147676NB-I00). Y.R. and A.O. acknowledge support from the ONRG (Award No. N62909-22-1-2031). This research was conducted within the framework of the Transnational Common Laboratories (LTC) Aquitaine-Euskadi Network in Green Concrete and Cement-based Materials and TRANS-LIGHT.

**Data Availability Statement**

Data sharing is not applicable to this article as no new data were created or analyzed in this study.

**Author contribution**

**Jehyeok Ryu:** Conceptualization (equal); Formal analysis (lead); Investigation (equal); Visualization (lead); Writing – original draft (equal); Writing – review & editing (equal). **Victor Krivenkov:** Conceptualization (equal); Investigation (equal); Supervision (equal); Visualization (supporting); Writing – original draft (equal); Writing – review & editing (equal). **Adam Olejniczak:** Conceptualization (supporting); Investigation (supporting); Visualization (supporting); Writing – original draft (equal). **Alexey Y. Nikitin:** Funding acquisition (equal); Supervision (equal); Writing – review & editing (equal). **Yury Rakovich:** Funding acquisition (equal); Supervision (equal); Writing – review & editing (equal).